\documentclass{article}

\usepackage{amsmath}
\usepackage{graphicx}

\usepackage[utf8]{inputenc}
\usepackage[english]{babel}
\usepackage[
backend=biber,
style=alphabetic,
sorting=ynt
]{biblatex}
\usepackage[dvipsnames]{xcolor}

\usepackage[pdf]{graphviz}
\usepackage{tikz}
\usepackage{tikz-cd}
\usepackage{multicol}
\usepackage{authblk}
\usepackage{mathtools}
\usepackage{caption}
\usepackage{pgfplots}
\usepackage{hyperref}
\usepackage{subfig}
\usepackage{subfig}
\pgfplotsset{width=6.5cm, compat=1.18}  

\usepackage{enumitem}
\usepackage{smartdiagram}

\usepackage{float}
\usepackage{braket}

\begin{document}
\title{Design Verification of the Quantum Control Stack}

\author{Seyed Amir Alavi}
\author{Samin Ishtiaq}
\author{Nick Johnson}
\author{Rojalin Mishra}
\author{Dwaraka O N} 
\author{Asher Pearl}
\author{Jan Snoeijs \footnote{Authors listed in alphabetical order.}}
\affil{Riverlane Research Ltd, Cambridge, UK.}

\date{}

\maketitle

\begin{abstract}
This paper describes the verification of the classical software and hardware stack that is used to control cold atom- and superconducting-based quantum computing hardware. The paper serves both as an introduction to quantum computing and to how classical device verification techniques can be employed there. Two main challenges in building a quantum control stack are generating precise deterministic-timing operations at the edge and scaled-out processing in the middle layer. Both challenges are to do with a certain kind of functional performance correctness. And, as usual, the design lives under tight power, memory and latency constraints. The quantum control stack is a complex interaction of algorithms, software runtimes and digital hardware. We take inspiration from modern software approaches to engineering, such as continuous integration and hardware automation, to quickly ship experimental features to customers in the field.
\end{abstract}

\section{Introduction}
The Church-Turing thesis states that all computers perform the same up to a polynomial factor in execution speed. Quantum computers (QC) are the only devices known to violate this thesis, promising a quantum speedup over classical computers that may permit solutions to hitherto intractable problems in science and mathematics. Quantum mechanics is a stranger place than the everyday Newtonian world of state machines. In classical computing, the unit of information, the bit, can be in one of the two states, 0 or 1. The quantum equivalent to the bit is the qubit, representing superposition, $\alpha \ket{0} + \beta \ket{1}$, where $\alpha$ and $\beta$ are complex numbers. Consequently, information stored in a quantum computer grows exponentially with every additional qubit. Moreover, the interconnected nature of the state space for two or more qubits allows entangled states to be generated that are impossible to produce with classical bits. These characteristics are expected to enable quantum computers to tackle problems that are computationally hard for classical algorithms, such as factoring large numbers or simulating quantum interactions with high accuracy. Famous example applications are breaking the RSA public-private key system security and cracking Bitcoin encryption, or modelling chemical processes of industrial relevance such as catalysis in the Haber process using the iron-molybdenum cofactor (FeMo-co) \cite{Webber}. 

Much research is underway on building the first quantum computers. A diverse range of qubit technologies exist, and several approaches have demonstrated high-fidelity quantum operations with the potential to scale up to large numbers of qubits. Some examples are trapped ions confined within electric fields, and superconducting circuits held at extremely low temperatures (Section~\ref{sec:hw}). However, qubits are extremely sensitive both to the environment and to the signals used to control them. These control signals, consisting of highly-tuned laser and radiofrequency (RF) electronic pulses, require high-precision manipulation of the amplitude, frequency and phase of the waveforms (Section~\ref{sec:controlsystem}). The more accurately we can manipulate these parameters, the better we can control the quantum state. If we were only talking to digital designers, we might conspiratorially admit that, while better qubits are the foundation of quantum computing, better digital control of the qubits is a big part of making quantum computing useful. (Just as a software (SW) person would say that SW is what makes classical hardware (HW) useful!) 

To overcome the fragility of individual qubits, quantum error correction (QEC) is implemented by forming distributed ‘logical’ qubits. Each logical qubit is constructed by encoding the quantum state over multiple physical ‘data’ qubits. In practice, these physical qubits are, for example, individual trapped ions. Since direct measurements of the data qubits are forbidden due to the destructive effect on the quantum state, additional so-called ‘auxiliary’ qubits are introduced which can interact with the data qubits to yield information about their underlying states via parity checks. In this way, errors such as bit flips can be detected (to within a certain confidence). The surface code is one of the leading encoding schemes for fault tolerant (FT) QC \cite{Litinski}. The job of a quantum error correction algorithm is to accurately infer the states of the data qubits from parity check measurements made only on the auxiliaries, and then, once armed with this information, to trigger operations on the data qubits that either preserve or reset their states. Owing to short quantum coherence lifetimes (typically ms for ion qubits, $\mu$s for superconducting qubits), this procedure places challenging demands on the timing accuracy and feedback responsiveness of the control system.

The control stack operates under tight power, memory, and latency constraints. But, because of the requirement for accurate RF signals, there is also an important functional performance correctness criteria in design verification (Section~\ref{sec:verification}). These would be challenging problems in any area but, as the quantum computing industry moves from university labs to deep-tech scale-ups, requirements change very frequently. Design verification now has the role to quickly qualify the current design in a situation where there are many moving parts of the full SW and HW stack. We take a more agile, SW-inspired approach to full-stack device verification and shipping (Section~\ref{fig:CI}). 

\section{Related Work}
\label{sec:related}
Google and ETH detail how the surface code is implemented on superconducting systems \cite{Google, Krinner}. Previous work at Delft is experimental \cite{Krinner} and has filled-out some of the HW micro-architecture gaps to implement surface code-based fault-tolerant computing \cite{Fu}. The papers we cite here are predominantly from large, well-organized research labs where the customer and the design engineering are the same team. These organizations have single-mission funding and a fixed set of internal-only requirements. The papers describe gigantic feats of scientific achievement, but ones that cannot easily be replicated outside that organization. The papers do not describe replicable, automated design verification. 

\section{Quantum Hardware}
\label{sec:hw}
A proliferation of emerging quantum computing hardware platforms exist, including photonics, silicon spin qubits, and NV centres in diamond. In this paper we focus our attention on trapped-ion and superconducting qubits. In a trapped-ion architecture, the qubits are single atomic ions carrying an overall positive electric charge. The ion is trapped by a time-varying electric field that is generated by the ion-trap device itself – usually microfabricated onto a chip by depositing metal electrode layers onto an insulating substrate like silicon dioxide. Typically, the ion trap chip is the size of a postage stamp and the ion is held above the surface at a height of around 100 $\mu$m – approximately the width of a human hair. Laser cooling techniques are used to bring the effective temperature of the ion to a few $\mu$K. The qubit states $\ket{0}$ and $\ket{1}$ are chosen to be a ground state and an excited state of the electronic energy levels within the ion. Control over the quantum state is achieved using pulsed laser- and microwave-fields with modulated amplitude, frequency and phase.

Superconducting qubits combine the operating principles of resonant LC circuits with the quantum behaviour of a Josephson junction. Quantum circuits containing many superconducting qubits can be fabricated onto microchips that operate at mK temperatures within a 3He/4He dilution refrigerator. This ‘dil fridge’ is a much-photographed device and looks like a gold-plated chandelier. A mix of microwave frequency, RF and DC control signals are used to manipulate the quantum states of superconducting qubits. IBM and Google have focused to develop this technology as part of their quantum computing research efforts.

\section{A Control System}
\label{sec:controlsystem}
Previously, we said that quantum states are controlled using pulsed laser or electronic signals with modulated amplitude, frequency and phase. These signals have to be transmitted with low-latency and high timing-accuracy. To begin to do a basic operation like initialising a qubit into a particular state, we need to sequence a set of pulses to run with at least ns timing accuracy. The definition of this sequence can be written in a programming language like OpenQASM’s OpenPulse~\cite{openpulse}. The definition is then compiled into instructions sent to a hardware execution unit that controls digital and RF signal generators. At runtime, this execution unit calculates and transmits the signals with clock accuracy. Timing-determinism is implemented by the execution unit queuing instructions up for the signal generators and firing them from the queue at the correct time. In practice, each execution unit controls 10s of such signal generators, and there will be 10s, 100s, even 1000s (for FT QC) of these execution units. Early in the design cycle, a set of FPGAs is the obvious choice of lego-bricks to build this system. The set of FPGAs are coordinated and synchronised at configuration time via software, and at run time via hardware signals. We know that cost pressures and increased specialisation will cause a move to more dedicated custom hardware, at the price of reduced flexibility of configuration/design re-spins. Our control system is currently implemented on Xilinx RFSoC ZCU111 FPGAs.

These pulse sequences are the lowest level of control. A set of pulses can be arranged into higher-level operations like qubit initialisation (init0, init1) and physical qubit gates (X, H). A quantum hardware system often requires early-morning calibration to discover the most accurate definitions of these higher-level operations. For example, there may be drifts in parameters such as frequency or amplitude over time, requiring compensation or re-calibration. Typical calibration workflows in a QC lab will involve scans over a range of control frequencies or power levels, to pinpoint the optimum qubit response. One of the most common scanning procedures is used to drive the qubit between the $\ket{0}$ state and the $\ket{1}$ state (and back again), to identify a characteristic oscillation period for the qubit. These are known as Rabi oscillations and the procedure is useful for calibrating the pulses which form gates when applied to the qubit.

Once we have pulse-level definitions of physical gates, then these can be abstracted again to define a physical circuit to implement error correction. QEC is just a certain quantum circuit run over and over, which involves gates and measurements. In practice, the control system for FT QC will also include modules on logical state management, noise modelling, etc.

\section{Control System Verification}
\label{sec:verification}
We have built three flows to implement completeness and redundancy for full-stack verification. These flows use all the classical verification techniques we have in our armour to ensure full-system functional and performance (timing determinate) correctness. The first flow (Section~\ref{sec:uvm}) is a very traditional UVM-based verification. The second flow (Section~\ref{sec:systemc}) is a SystemC model of the top-level combined with parts of the SW stack. The third flow (Section~\ref{sec:hwlab}) is a pretty faithful lab mock-up of the actual in-field SW and HW. 

We have a continuous integration (CI) system (Section~\ref{sec:ci}) that kicks off tests on both source and, transitively, on upstream/dependent repos. The flow aims to ensure that a good version of the entire stack is always ready for release. This system is designed to be automatic, discovering a “good frontier” that is always ready for release across the set of source git repos.

\subsection{Top/Block level UVM Verification}
\label{sec:uvm}
The first flow is a very traditional top- and block-level UVM-based verification. This flow tests the custom HW using SystemVerilog simulation and formal tools. The flow is cycle-accurate and allows full debugging of the design but is slower than real hardware and can’t be used to do full-stack verification. The architecture of the control system has both DSP and Digital blocks. The critical blocks have been prototyped as functional SystemC or Python implementations and are used as golden models.

To verify a single block of the quantum control stack using UVM, constrained randomized data is driven from the top-level using multiple sequences in parallel, which helps in making sure that data that flows throughout the stack is always consistent. The integrity of data generated is checked in UVM using scoreboards and reference models. Scoreboards in the test bench are used for all the digital blocks of the architecture. The data generated by the testbench and RTL are compared for correctness.

The DSP reference models define the shape and spectral purity of the generated sinusoid. The models have reals as their native number representation. One of the challenges has been keeping tabs on the number representation formats in comparing the reference model to the RTL implementation. In our design flow for one of the signal generation blocks, the Python reference model uses floating point numbers, while the RTL implementation uses normalized fixed point numbers. There are a variety of different formats in the data processing chain, and we have convertors between them to verify the RTL. We have needed to strongly specify the number of fractional digits, for instance; and to check how different tools store and transmit fixed/floating point numbers. 
One of the flows, illustrated in Figure~\ref{fig:refmodel} below, also includes several conversion utilities: shown here are a fixed-to-floating converter, and a convertor to represent the config to the DUT’s config RAMs. 

\begin{figure}[h]
    \centering
    \includegraphics[scale=0.75]{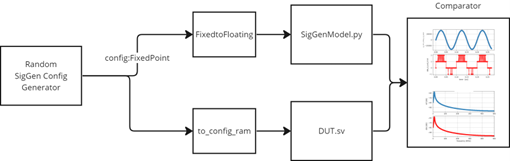}
    \caption{Using a reference model to check a DSP implementation.}
    \label{fig:refmodel}
\end{figure}

The expected and actual output digital samples are compared and checked to see that they are within the allowed threshold. In Figure~\ref{fig:refmodel} the first two Comparator graphs are the red/actual and the blue/expected pulses (they are very close, and the blue expected completely overlaps the red actual), and the error bars. The last two Comparator graphs are the blue/expected and red/actual spectral graphs for the output. We applied Blackman windowing to the expected and actual digital data. As the graphs show, the tones both appear at driven frequency and with no unexpected harmonics within the frequency band.
Only checking the data integrity would not be sufficient for any design. The UVM testbench makes sure that the design is cycle accurate. Cycle accuracy in the testbench is checked by assertions to measure latency between inputs and outputs of each block. In Figure~\ref{fig:siggen} below, some validity assertions are shown for one of the signal generation blocks. 

\begin{figure}[h]
    \centering
    \includegraphics[scale=00.75]{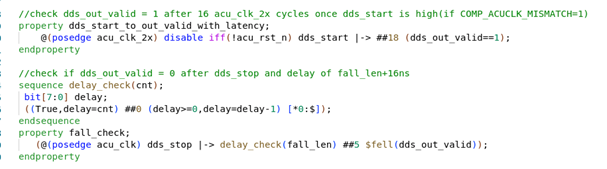}
    \caption{SignalGen block valid timing assertions.}
    \label{fig:siggen}
\end{figure}

We use formal verification (FV) extensively to verify the interconnects of our architecture.

The completeness of verification is measured by checking the functional and code coverage. Functional coverage is checked with cover-points and coverage bins programmed for each individual block. Code coverage is checked by running multiple regressions and merging all data collected in each run. We use formal verification to also speed up functional and code coverage, and for extracting cover-points and eliminating dead code.

\subsection{Top-level SystemC Model}
\label{sec:systemc}
In this section, we describe the second verification flow. This consists of a SystemC~\cite{systemc} model of the top-level --- with some non-critical blocks being replaced by functional models --- combined with a subset of the SW stack. Including parts of the SW stack means that we are now running more of the full-stack system in simulation and can run user-level programs. Even though it is clock-level accurate, we get a factor improvement over the HW simulators, enabling a faster turnaround for early bug discovery. 

\begin{figure}[h]
    \centering
    \includegraphics[scale=00.75]{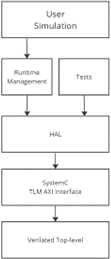}
    \caption{SystemC/C++ simulator SW stack.}
    \label{fig:systemc}
\end{figure}

The SW stack is shown in Figure~\ref{fig:systemc}. The bottom layer is the Verilated top-level HW block in a SystemC wrapper. This layer is connected to a SystemC transaction layer modelling (TLM) interface to execute Advanced eXtensible Interface (AXI) bus transactions to communicate with the blocks. From this layer, the same hardware abstraction layer (HAL) that is used in production can be used to interface with the HW blocks. This level of abstraction brings a lot of benefits such as testing production SW and easier maintenance of verification, control and runtime code. At the highest level, the simulator exposes both a SystemC and control API for execution of the simulation and post-processing the log the simulation.

A Verilated model is very fast compared to running the design using the usual SystemVerilog simulators due to the speed gain from native compiled C++. To further improve the simulation speed, we have enabled multithreading in the verilated model. This boosts the simulation speed by the number of available processor cores on the machine. Although this feature is very useful, it has a major limitation: the number of threads have to be fixed during Verilator compilation (converting SystemVerilog to C++) time, which doesn’t support optimizing the number of threads at runtime. The user API interface for the simulator provides basic operational interface, such as starting, stopping, signals to be recorded, initial configuration of the block, etc. The API interacts with the runtime management system to synchronize operations between different HW blocks and SW components of the HAL.

\subsection{On-target automated tests in HW Lab}
\label{sec:hwlab}
\begin{figure}[h]
    \centering
    \includegraphics[scale=00.75]{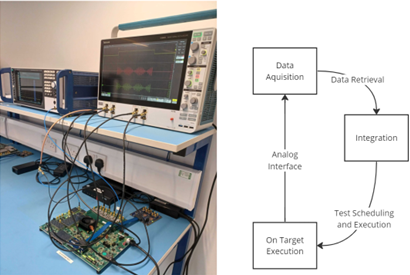}
    \caption{On-target HW automated tests setup.}
    \label{fig:hwlab}
\end{figure}

Verifying the functionality of the control system components on the target HW is a crucial step to confirm actual expected behaviour of the simulation results. In this step, all of the SW and HW components are deployed on the target. We have custom tooling that runs integration tests automatically on the target: it executes the test cases on the target, and captures the results on the physical ports using the allocated data acquisition device, in this case a network-controllable oscilloscope. This setup is shown in Figure~\ref{fig:hwlab}

There are limitations to be considered when designing the HW automated tests, which are mainly the constraints on the data acquisition device such as the number of available ports, and available memory/bandwidth for data retrieval. This leads to a trade-off between resolution and length of sampling. Another complexity worth mentioning here is the dynamic nature of the HW blocks integrated into the FPGAs, which requires continuous change in the HW connections and data capturing configuration. Although this step gets complicated in practice, it provides valuable insight on performance of the components leading to reduced risk of bugs or other issues after product releases.

\subsection{Full-stack CI}
\label{sec:ci}
The source code for the control system is stored over nine git repos. Each repo is logically contained in the sense that all the FPGA SystemVerilog source is stored in one repo, all the low-level embedded software is stored in one repo, etc. Each repo has it’s own set of testing and verification artifacts and flows. When commits on a feature branch are ready to be merged into the repo’s main branch, a full CI runs on the new branch. 

\begin{figure}[h]
    \centering
    \includegraphics[scale=00.75]{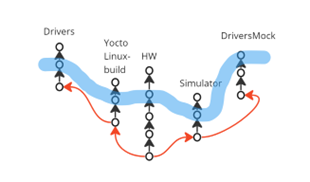}
    \caption{CI Good Frontier.}
    \label{fig:CI}
\end{figure}

This isolated set up ignores the fact that the repos are the source of artifacts that build upon each other. This is true in both the source sense (a C++ module dependent upon the source C++ module in another repo) and in the binary sense (a source Python module dependent upon a C++ compiled binary) too. In Figure~\ref{fig:CI} above, the history of each repo is shown going up; the red arrows indicate a dependency (via git modules, package managers or manual installs). In our experience, interesting bugs occur when several non-trivial repos’ artifacts are run together. This makes sense intuitively: we think we have completely understood and tested a module, but its interactions with other modules in a larger sub-system bring out all the hidden assumptions we have on the module’s interfaces. So, the earlier “full stack” integration can be done, the sooner these critical late bugs can be found, and the quicker the turnaround for new features.

The way we run full-stack integration testing is as follows. Every repo has a CI-passing branch that points at the latest commit close to main (could be main) at which all the repos work well together. The CI-passing pointers form a “good frontier”, illustrated by the blue border in  Figure 5 above, that is always ready for release across all nine git repos. If CI passes on repo A’s main, then A’s CI-passing branch is updated to main. All submodules automatically track head of CI-passing which is the latest point that all integration tests for all submodules passed.

The working model here is that an engineer checks out repo A’s main. All of A’s submodules track CI-passing, and are automatically checked out at a working commit that works with the head of main. The engineer makes a new feat-br off main, does some work and makes a PR. On creation of the PR: (1) CI on repo A tries to test A’s feat-br with submodule B’s CI-passing; (2) CI on repo A tries to update submodule B’s ci-passing to head of main and tries to run integration tests between A on new-branch and B. At least one of (1) or (2) should pass. If both fail then there’s a problem, and the PR shouldn’t be merged. If (2) fails but (1) passes, that serves as a good ongoing indication that work is needed to fix integration between A and B without immediately breaking A.

After the merge of the PR, full-stack CI checks out all submodules at head of main, and runs integration tests performing any updates to CI-passing on submodules if all integration tests pass. If any fail, CI-passing on all submodules are kept at the older commits that worked. This ensures that if A is itself a submodule of some other repo R, we don’t end up in a situation where we’ve updated the ci-passing on B, but the head of main on A doesn’t work with R, and so CI-passing on A doesn’t change, which would cause CI-passing to no longer reflect a set of commits that all work together.

This model ensures that CI-passing is close to or at the head of main for all the submodules (if it isn’t then that’s indicative of a larger problem that needs to be fixed before release). By definition, CI-passing is “the latest set of commits which work together correctly”, so should is always releasable. We check out CI-passing for our release commits. Using this methodology, we are able to release every two-week sprint.

\section{Further Work}
\label{sec:further}
We have described the design verification flow for a quantum control system. We see this as the 1st generation of control systems. There are two large design trajectories in our current work. First, scaling out the system to 10s, 100s, 1000s of qubits in anticipation of fault tolerant quantum computing. We are working to analyse and model functional, performance, protocol, memory, etc requirements of scaled-out control system for different quantum hardware. Second, portability over this different quantum hardware. How can we keep the design configurable over different target systems and hardware, and how can we verify over these configurations? It is quite likely that each generation might be radically different from the previous one. So, we are working through the implications for sub-system and full-system design verification. In typical HW development, old source (not necessarily module) code is salvaged and re-used. But modern agile ways of working provide good guides to more sure ways of qualifying large-scale complex design. 

\medskip 

\printbibliography

\end{document}